\documentstyle[prb,aps,preprint,eqsecnum,epsf]{revtex}

\makeatletter
\def\preprint#1{%
\def\@preprint{\noindent\hfill\hbox{#1}\vskip 10pt}%
}
\makeatother

\begin{document}
\draft
\preprint{ITP-UH-01/97}

\title{Variational states for the spin-Peierls system}
      
\author{Holger Frahm\cite{email_hf} and John Schliemann\cite{addr_js}}
\address{Institut f\"ur Theoretische Physik, Universit\"at Hannover
	D-30167~Hannover, Germany}
\date{January 1997}
\maketitle
\begin{abstract}
We introduce a family of Jastrow pair product states for quasi
one-dimensional spin systems.
Depending on a parameter they interpolate between the resonating valence
bond ground state of the Haldane--Shastry model describing a spin liquid
and the (dimerized) valence bond solid ground states of the Majumdar--Ghosh
spin chain.  These states are found to form an excellent basis for
variational studies of Heisenberg chains with next nearest neighbour
interaction and bond alternation as realized in the spin-Peierls system
{CuGeO$_3$}.
\end{abstract}


\section{Introduction}
Following the recent discovery of a spin-Peierls transition in the
inorganic compound {CuGeO$_3$} \cite{hase:93} there has been growing
theoretical interest in this instability of one-dimensional spin chains.
It has been proposed in Refs.~\onlinecite{lorx:94,cace:95} that frustrating
next nearest neighbour (nnn) interactions in addition to an explicitely
broken translational invariance due to lattice dimerization is necessary to
obtain a consistent description of the experimental data. This leads to the
following spin-${1\over2}$ Hamiltonian
\begin{equation}
   {\cal H} = \sum_{j=0}^{N-1} \left((1 + (-1)^j \delta) 
	{\mathbf S}_j\cdot {\mathbf S}_{j+1}
	+ \alpha {\mathbf S}_j\cdot {\mathbf S}_{j+2}\right)\ .
\label{hamil}
\end{equation}
For $\delta=0$, the model is invariant under translations by one lattice
site.  This case has been investigated in detail and includes the nearest
neighbour Heisenberg chain, $\alpha=0$, where the complete spectrum can be
obtained by means of the Bethe Ansatz\cite{bethe:31}: the ground state is
that of a spin liquid, has a vanishing spin gap and algebraically decaying
correlations at $T=0$.  For $\delta=0$, $\alpha={1\over2}$ the Hamiltonian
(\ref{hamil}) becomes that of the Majumdar--Ghosh (MG) model\cite{magh:69}:
here the system has a gap \cite{AKLT} and the exact ground state is known
to be a product of nearest neighbour singlet pairs showing a twofold
degeneracy
\begin{equation}
  |\psi_{VB}^{(1)}\rangle \propto [0\,;1]\cdots[(N\!-2)\,;(N\!-1)]\ ,
\qquad
  |\psi_{VB}^{(2)}\rangle \propto [1\,;2]\cdots[(N\!-1)\,;0]
  \label{VB}
\end{equation}
($[ a\,;b]=\frac{1}{\sqrt2} (|\uparrow\rangle_a |\downarrow\rangle_b
-|\downarrow\rangle_a|\uparrow\rangle_b)$ denotes the singlet state formed
by the spins on sites $a$ and $b$).
For intermediate values of the nnn interaction the model (\ref{hamil})
remains gapless for $\alpha<\alpha_c\approx 0.2411$
\cite{affx:89,okno:92,cace:95} with its low energy sector described by an
effective level $k=1$ SU(2) Wess-Zumino-Witten (WZW) conformal field
theory.  Increasing $\alpha$ beyond the `conformal point' $\alpha_c$ the
nnn coupling becomes marginally relevant, producing an exponentially small
gap $\Delta\propto\exp(\hbox{const.}/(\alpha-\alpha_c))$.  In this phase
the system is spontaneously dimerized.  The properties of the system for
$\alpha>1$ have recently been discussed in Ref.~\onlinecite{whaf:96}.
Extending the discussion to general couplings in the $\alpha$--$\delta$
plane the system has gap above a dimerized ground state for any nonzero
$\delta$,\cite{DimerGS} with the valence bond ground states (\ref{VB}) on
the line $2\alpha+\delta=1$ \cite{shsu:81}.  Going to larger $\delta$ the
Hamiltonian (\ref{hamil}) corresponds to a ladder system of two coupled
Heisenberg chains \cite{Ladders}.
In addition to numerical studies, several mean-field theories have been
proposed for the Hamiltonian (\ref{hamil}) \cite{MeanField}.
In this framework physical properties have been calculated giving
reasonable agreement with the experimental data available for energy gaps,
Raman spectra and the susceptibility of {CuGeO$_3$} and hence further
support the Hamiltonian (\ref{hamil}) as a model for this substance.

In this paper we propose a family of variational states for the ground
state and low lying triplet of the Hamiltonian (\ref{hamil}).  While it
will not be possible within this variational approach to compute the
thermodynamical quantities mentioned above, it turns out that the states
proposed are excellent approximations to the true ground state of the
system throughout the parameter region of interest here, namely
$2\alpha+\delta\le1$.  Hence they can be used to obtain very good
variational bounds on energies and may provide a better understanding of
the role of quantum fluctuations in these systems.

In fact, variational states have already been applied successfully to some
of the systems mentioned above: the Jastrow pair product wave function
\begin{equation}
  \vert \psi^{N,M,J}_0 \rangle =
    \sum_{{n_1,\ldots,n_M}} \psi(\{n_i\}) \prod_{i=1}^M S_{n_i}^-
    |\uparrow\cdots\uparrow\rangle\ , \qquad
  \psi(\{n_i\}) = \prod_{i=1}^M g(n_i) \prod_{i<j} d(n_i-n_j)^2
\label{jas_hs}
\end{equation}
with $d(n)=\sin(\pi n/N)$, $g(n)\propto\exp(2\pi i(J/N) n)$ and $M=J=N/2$
has been found to reproduce the ground state energy of the nearest
neighbour Heisenberg chain of $N$ sites with remarkable accuracy
\cite{kahf:82}.
Furthermore, (\ref{jas_hs}) captures the essence of the spin spin
correlations in this system.  States of the form (\ref{jas_hs}) span a
large part of the Hilbert space including the ground state of the Haldane
Shastry (HS) spin chain with long range exchange interactions
$J_{kl}\propto 1/\sin^2(\pi(k-l)/N)$ \cite{HaldaneShastry}.  The spectrum
of this model gives a representation of the $k=1$ SU(2) WZW conformal field
theory \cite{hald:91} (note that the next nearest neighbour interaction in
the HS model is very close to $\alpha_c$).

The spin Peierls system and ladder like models on the other hand have been
studied using short range RVB states as ground states and soliton states
interpolating between the two singlet bond configurations (\ref{VB}) for
the excitations \cite{ValenceBonds}.

In the following section we introduce variational states depending two free
parameters and prove that they are eigenstates of the total spin of the
system.  In Section~\ref{sec:cases} we show that these states contain the
valence bond states (\ref{VB}) and the Jastrow state (\ref{jas_hs}) as
certain limits.  Their relation to Gutzwiller projected states of
spin-${1\over2}$ fermions is discussed in \ref{sec:gutz}.  Finally, we
apply the variational states to the spin-Peierls system (\ref{hamil}) in
Section \ref{sec:app}.

%
%
\section{The variational states}
\subsection{Construction}
\label{sec:states}
Let $C_1$ and $C_2$ be two parallels with polar angles $\theta$ and
$\pi-\theta$ respectively on the two-dimensional sphere $S^2$.  For even
$N$ we choose $N/2$ equidistant points labeled by even numbers
$0,2,4,\ldots,(N-2)$ on $C_1$ and similarly $N/2$ points labeled by odd
numbers $1,3,5,\ldots,(N-1)$ on $C_2$ such that their azimuthal angles
satisfy
\begin{equation}
\frac{2\pi}{N}(1-2\rho)=\varphi_{2i+1}-\varphi_{2i}
\end{equation}
for $i\in \{0,\dots,(N-2)/2\}$.  This defines a lattice on $S^2$
characterized by the parameters $\theta$ and $\rho$ consisting of two
identical linear sublattices and obeying periodic boundary conditions in
the azimuthal direction.  The lattice points $\vec{a}$ can be described in
terms of polar coordinates on the sphere, i.e.\ $\vec{a}= (\cos\varphi_a
\sin\theta_a, \sin\varphi_a \sin\theta_a,\cos\theta_a)$. Alternatively,
they may be parametrized by spinor components
$u_{a}=\cos(\theta_{a}/2)\exp(\imath\varphi_{a}/2)$, $v_{a}=\sin(\theta_{a}/2)
\exp(-\imath\varphi_{a}/2)$.  Given two points $a,b\in S^2$ we construct the
spinor product
\begin{equation}
   d(a,b)=(u_{a}v_{b}-u_{b}v_{a})\ .
\label{spinorp}
\end{equation}
that has the following properties:
\begin{eqnarray}
|d(a,b)|^2&=&\frac{1}{4}\|(a-b)\|^{2}\ ,
\nonumber \\
   \arg(d(a,b))&=&\tan^{-1}\left(\frac{\sin((\theta_{a}+\theta_{b})/2)
   \sin((\varphi_{a}-\varphi_{b})/2)}
   {\sin((\theta_{a}-\theta_{b})/2)\cos((\varphi_{a}-\varphi_{b})/2)}\right)\ ,
\end{eqnarray}
where $\|\cdot\|$ stands for the euclidean norm in ${\rm R}^3$.

Now let each lattice point carry a spin ${\mathbf S}_{n}$ with
$S_{n}=\frac{1}{2}$. In the Hilbert space of these $N$ spins we consider
the following states ($d$ given by (\ref{spinorp})):
\begin{equation}
  |\psi^{N,M,J}(\theta,\rho)\rangle \propto
  \sum_{|\{n_{i}\}|=M}\prod_{i}e^{\imath
  \frac{2\pi}{N}Jn_{i}}
  \left(\prod_{j\atop i<j}d^2(n_{i},n_{j})\right)\quad\!\!\!\!
  S_{n_{i}}^{-}|\uparrow\cdots\uparrow\rangle\ .
\label{DJS}
\end{equation}
Here $|\uparrow\cdots\uparrow\rangle$ is the ferromagnetically ordered
state with all spins parallel. The sum extends over all possibilities to
select $M$ out of $N$ lattice sites and invert their spins. The sets of the
selected sites are denoted by $\{n_{i}\}$. Each of those spin-product
states contributes with an amplitude that consists of a product of
single-site phasefactors and a product of Jastrow-like two-site factors,
depending on the underlying lattice, namely the parameters $\theta$ and
$\rho$. The spinor components on the $n$th lattice site may be rewritten up
to a common factor as $u_{n}\propto\exp(+\imath
((\pi/N)(n+(-1)^{n}w))$, $v_{n}\propto\exp(-\imath
((\pi/N)(n+(-1)^{n}w))$ with $w=\rho+\imath \kappa$ and
$\kappa=-(N/2\pi)\ln(\cot(\theta/2))$. 
Hence, the states defined in (\ref{DJS}) depend analytically on a complex 
parameter expressing a dimerized structure of the lattice and will be referred 
to as {\it dimerized Jastrow states} (DJS).  
Periodic boundary conditions require the parameter $J$ to be taken integer.  
Requiring the DJS to have a definite total spin restricts $J$ even further:\\
Clearly, the (\ref{DJS}) are eigenstates of the $z$-component of the total
spin with eigenvalue $N/2-M$.  To show that they are eigenstates of the
total spin we rewrite ${\mathbf S}^2$ as $\sum{\mathbf S}_{n}^2 +\sum_{n<m}
(S_{n}^{+}S_{m}^{-}+S_{n}^{-}S_{m}^{+}+2S_{n}^{z}S_{m}^{z})$.  Obviously
$\sum_{n}{\mathbf S}_{n}^2\,|\psi^{N,M,J}(\theta,\rho)\rangle=\frac{3}{4}N\,
|\psi^{N,M,J}(\theta,\rho)\rangle$ and
\begin{eqnarray}
   &&\sum_{n<m}2S_{n}^{z}S_{m}^{z}\,|\psi^{N,M,J}(\theta,\rho)\rangle =
\nonumber\\
   &&\qquad=
     \frac{1}{4}\left(M(M\!-1)+(N\!-M)(N\!-M\!-1)-2(N\!-M)M\right)\,
	|\psi^{N,M,J}(\theta,\rho)\rangle.
\label{SzSz}
\end{eqnarray}
For integer $J$ with $M\leq J\leq N-M$ this gives together with the results 
of appendix \ref{appa}:
\begin{equation}
   {\mathbf S}^2\,|\psi^{N,M,J}(\theta,\rho)\rangle
   =(\frac{N}{2}-M)(\frac{N}{2}-M+1)\,|\psi^{N,M,J}(\theta,\rho)\rangle\,.
\label{S^2a}
\end{equation}
Hence, under the above condition the state
$|\psi^{N,M,J}(\theta,\rho)\rangle$ is an SU(2) highest weight state with
total spin $S=N/2-M$ for arbitrary values of the parameters $\theta,\rho$.
In particular, choosing $M=J=N/2$ one obtains a singlet,
$M=N/2-1,\,J\in\{N/2-1,N/2,N/2+1\}$ gives three triplet states with
$S^{z}=1$ etc.
\subsection{Limiting cases}
\label{sec:cases}
From (\ref{spinorp}) one finds that the amplitudes of the wave functions
(\ref{DJS}) will be complex in general.  These 'chiral' spin states are
believed to arise in two dimensional spin systems due to frustrating
interactions \cite{wwz:89} (see also Section~\ref{sec:ground} below).
Choosing $w\in {\rm R}$, i.e. $\theta=\frac{\pi}{2}$, all lattice sites lie
on the equator of the sphere and the Jastrow-factors in (\ref{DJS}) become
all real: now the relative phases of the amplitudes are determined by the
one-site factors alone.  In this case, some of the states discussed in the
introduction can be obtained by properly choosing $\rho$:

\underline{\emph{(i)} $w=0$.}
Provided that $M-1\leq J\leq N-M+1$, the DJS
\begin{equation}
   |\psi_{HS}^{N,M,J}\rangle\,:=\,|\psi^{N,M,J}(\frac{\pi}{2},0)\rangle
\label{defHS}
\end{equation}
are eigenstates of the Haldane--Shastry model \cite{HaldaneShastry}
\begin{equation}
   H_{HS}=\sum_{n,m=0\atop n<m}^{N-1}\frac{1}{(\frac{N}{\pi}
   \sin(\frac{\pi}{N}(n-m)))^2}{\mathbf S}_{n}{\mathbf S}_{m}
\label{HSM}
\end{equation}
with the eigenvalues
\begin{equation}
   E_{N,M,J}=\frac{1}{3}(\frac{\pi}{N})^2
   (\frac{N}{8}(N^2-1)+M(M^2-1)-3MJ(N-J))
\label{EVHS}
\end{equation}
The ground state is given by the singlet, the first excitations by the
triplet with $M=N/2-1$,$J=N/2$.\\ 
In the thermodynamic limit the two spin correlation functions have been
evaluated exactly by Gebhard and Vollhardt \cite{GeVo}. For the singlet
state of (\ref{defHS}) they have obtained
\begin{equation}
   {\lim_{N\to\infty}}
   \langle\psi_{HS}^{N,\frac{N}{2},\frac{N}{2}}\,|\,
   {\mathbf S}_{k}{\mathbf S}_{k+n}\,|\,
   \psi_{HS}^{N,\frac{N}{2},\frac{N}{2}}\rangle
   =(-1)^{n}\frac{3}{4}\frac{{\rm Si}(\pi n)}{\pi n}
\label{CFHS}
\end{equation}
This result is based on an alternative formulation of the DJS that is
discussed below.

\begin{figure}
\caption{Overlap $|\langle\psi^{N,N/2,N/2}(\frac{\pi}{2},0)
|\psi^{N,N/2,N/2}(\frac{\pi}{2},\frac{1}{2})\rangle|^2$ between the ground
state of the Haldane--Shastry model and the valence-bond-state.
\label{ovl}}
\begin{center}
\epsfxsize=\textwidth
\epsffile{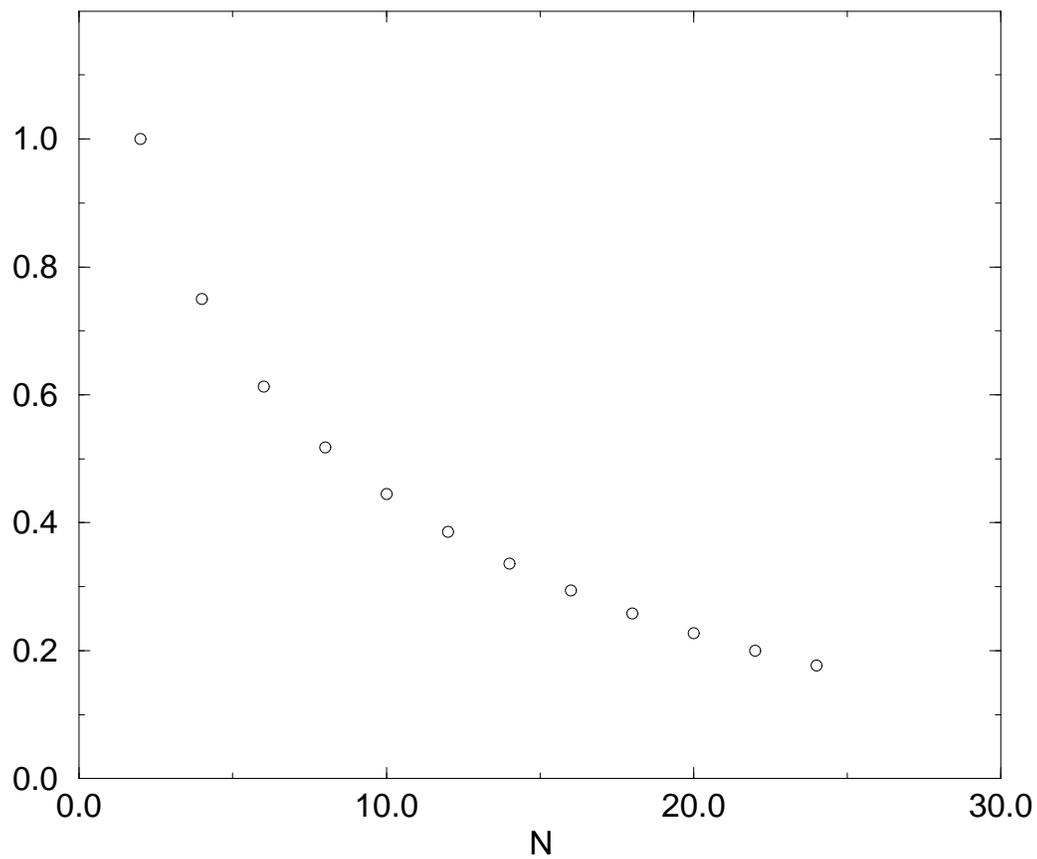}
\end{center}
\end{figure}

\underline{\emph{(ii)} $w=\frac{1}{2}$.}
Here an even-numbered lattice site coincides with the next higher site. It is
\begin{equation}
  |\,\psi^{N,\frac{N}{2},\frac{N}{2}}(\frac{\pi}{2},\frac{1}{2})\rangle
  \propto |\,\psi_{VB}^{(1)}\rangle\ .
  \label{VB2}
\end{equation}
The singlet of the DJS at $w=1/2$ is a valence-bond-state: next neighbours
are coupled to singlets.  (\ref{VB2}) is easily proved by calculating the
scalar product of both sides.  Similarly, we have
$|\,\psi^{N,\frac{N}{2},\frac{N}{2}}(\frac{\pi}{2},-\frac{1}{2})\rangle
\propto|\psi_{VB}^{(2)}\rangle$ and
\begin{equation}
|\,\psi^{N,\frac{N}{2}-1,\frac{N}{2}}(\frac{\pi}{2},\frac{1}{2})\rangle
\propto\sqrt{\frac{2}{N}}\sum_{k=0}^{\frac{N}{2}-1}
\lbrack0\,;1\rbrack\cdots|\uparrow\rangle_{2k}\!|\uparrow\rangle_{2k+1}\!
\cdots\lbrack(N\!-2)\,;(N\!-1)\rbrack
\end{equation}
DJS with higher total spin cannot be cast in a comparably simple form at
$w=1/2$ due to the more complicated structure of the remaining
amplitudes.\\
The two valence-bond singlets span the ground state space of the
Majumdar--Ghosh model \cite{magh:69}
\begin{equation}
  H_{MG}=\sum_{n=0}^{N-1}({\mathbf S}_{n}{\mathbf S}_{n+1}
         +\frac{1}{2}{\mathbf S}_{n}{\mathbf S}_{n+2})
\label{MGM}
\end{equation}
Recently, Nakano and Takahashi have generalized this hamiltonian to a
variety of models with interactions of arbitrary range that have the same
property \cite{nata:95}. In the thermodynamic limit these models have a
finite gap for excitations over the twofold degenerate ground state leading
to spin spin correlations that decay exponentially at large distances.
This is in contrast to the Haldane-Shastry model which has no gap and
according to (\ref{CFHS}) correlations decaying algebraically.\\
Hence the DJS with $\theta=\pi/2$ and $M=J=N/2$ interpolate between a
'resonating valence bond' singlet for $\rho=0$ and the nearest neighbour
'valence bond solid' described by (\ref{VB}) for $\rho={1\over2}$.  Despite
their essentially distinct properties these states have a remarkably large
overlap.  In Figure~\ref{ovl} the squares of the overlaps as computed
numerically are plotted for systems up to $N=24$.  At $N=24$ the
singlet-subspace has the dimension 208012, so that a square overlap of
$0.17$ should be regarded as quite large.\\
It would be interesting if one could find a hamiltonian interpolating
between one of the models given in Ref.~\onlinecite{nata:95} and the
Haldane--Shastry model, so that the singlet-DJS is {\it always} the exact
ground state.

\underline{\emph{(iii)} {$\theta\to0$}.}
For $\theta\to0$ and finite $\rho$ the chains $C_{1}$,$C_{2}$ are drawn to
the poles of the sphere.  In (\ref{DJS}) all amplitudes and consequently
the normalisation sum become zero. In this limit tensor products of
Haldane--Shastry-typed states arise. In the simplest case, the singlet-DJS,
one obtains for even $N/2$:
\begin{equation}
\lim_{\theta\to0}\,|\,\psi^{N,\frac{N}{2},\frac{N}{2}}(\theta,\rho)\rangle
\propto\,|\,\psi_{HS}^{\frac{N}{2},\frac{N}{4},\frac{N}{4}}\rangle
\otimes\,|\,\psi_{HS}^{\frac{N}{2},\frac{N}{4},\frac{N}{4}}\rangle\,,
\label{2HSb}
\end{equation}
The first factor in the tensor product refers to $C_{1}$, the second to
$C_{2}$. Other DJS can also be examined by expliciting their dependence on
the parameters $\theta$ and $\rho$.

\subsection{Relation to the Gutzwiller wave function}
\label{sec:gutz}
It is worthwhile noticing that the DJS can also be formulated analogous to the
Gutzwiller wave function \cite{gutz:63}.\\
Let $a_{n\sigma}^{+},a_{n\sigma}$ be canonical creation and annihilation
operators on the $n$th site for spin-$\frac{1}{2}$-particles with
$S_{n}^{z}=\sigma$ and $|\,0\,\rangle$ the vacuum of the system .\\
For e.g. $M=J=N/2$ one can use the following construction: With the
definition
\begin{equation}
b^{+}_{k\sigma}(w)=\frac{1}{\sqrt N}\sum_{n=0}^{N-1}
e^{\imath k(n+(-1)^{n}w)}a_{n\sigma}^{+}
\label{tr}
\end{equation}
we have
\begin{equation}
|\psi^{N,\frac{N}{2},\frac{N}{2}}(\theta,\rho)\rangle
\propto {\cal P}\,
b^{+}_{k_{0}\uparrow}(w)\cdots
b^{+}_{k_{\frac{N}{2}-1}\uparrow}(w)\,
b^{+}_{k_{0}\downarrow}(w)\cdots
b^{+}_{k_{\frac{N}{2}-1}\downarrow}(w)
\,|\,0\,\rangle\
\label{altdef}
\end{equation}
with $k_{n}=-\frac{\pi}{2}+\frac{\pi}{N}+\frac{2\pi}{N}n$ and ${\cal P}$
being the Gutzwiller projector excluding double occupancies, i.e ${\cal
P}=\prod_{n=0}^{N-1}(1-n_{n\uparrow}n_{n\downarrow})$,
$n_{n\sigma}=a_{n\sigma}^{+}a_{n\sigma}$. Eq.~(\ref{altdef}) can be proved
using similar arguments as in Refs.~\onlinecite{HaldaneShastry}.  So the
DJS are deformed Gutzwiller wave functions characterized by a complex
parameter $w$.  Analytical results concerning the two-site correlations
have been found for $w=0$ \cite{GeVo,MeVo}.  For non-zero $w$ the
dimerization leads to complications that cannot be resolved following these
methods.

An obvious generalisation is to construct wave functions of the above type
with arbitrary filling.  Starting from states without double occupancies on
the equidistant lattice as considered in Ref.~\onlinecite{tJJast} one can
introduce double occupancies and dimerization while keeping the Jastrow
form of the states \cite{schliemann:dipl}.  Again they can be constructed
to be heigest weight states of the total spin. Since empty and doubly
occupied lattice sites do not contribute to the spin, there are large
additional possibilities of varying the wave functions in the case of
general filling and strength of the Gutzwiller projection.
%
%
%
\section{Application to the spin-Peierls system}
\label{sec:app}
We now use the DJS as variational ansatz for the low lying states of the
model (\ref{hamil}).  As mentioned above the analytical methods of Refs.\
\onlinecite{GeVo,MeVo} cannot be applied to the dimerized system, hence the
results presented below were obtained by numerical evaluation of the
relevant matrix elements for system sizes up to $N=26$ lattice sites.
\subsection{Ground state properties}
\label{sec:ground}
\begin{figure}
\caption{Variational ground state energy per spin at $\delta=0$ as a
function of $\alpha$.
\label{gs}}
\begin{center}
\epsfxsize=\textwidth
\epsffile{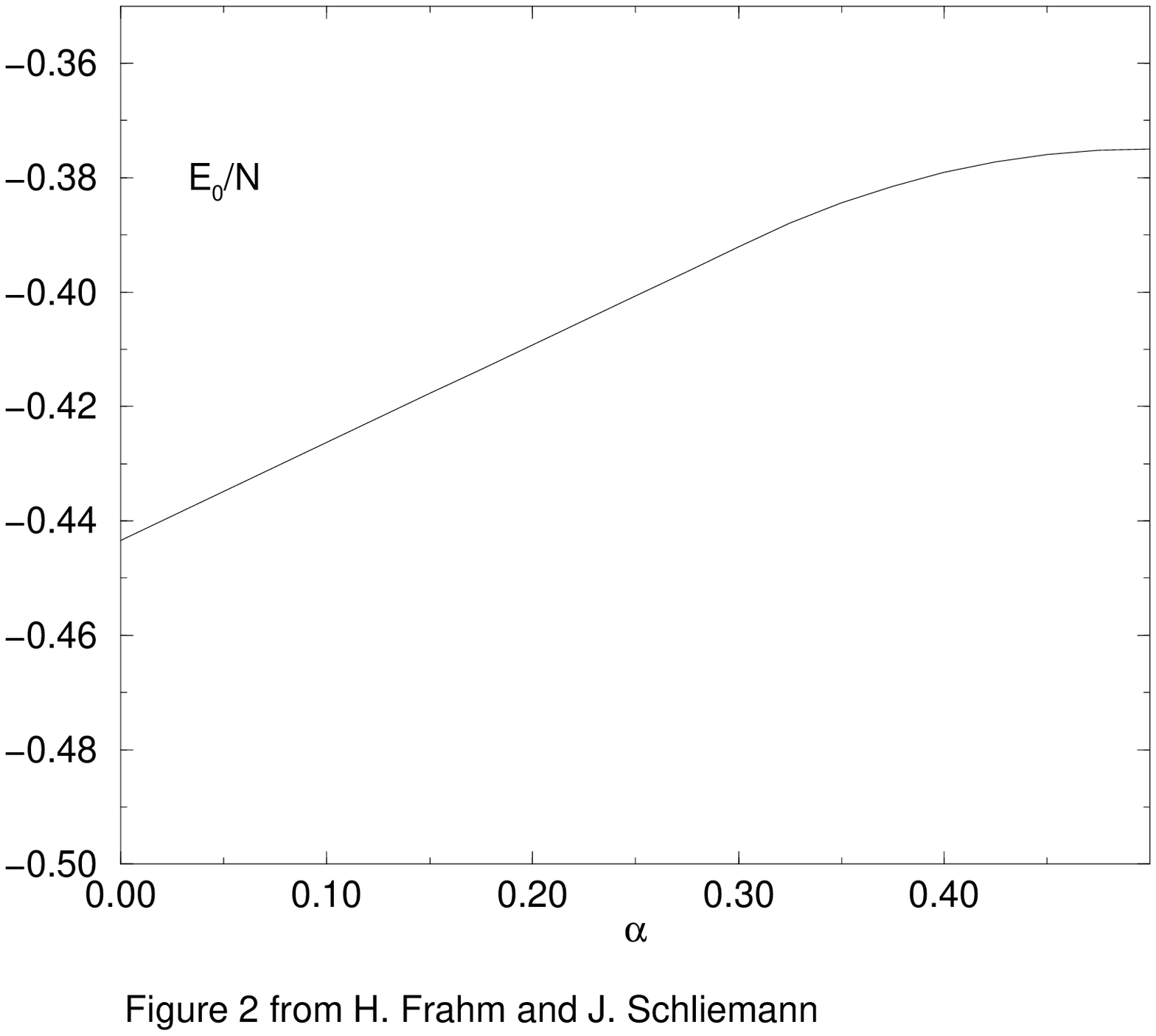}
\end{center}
\end{figure}

\begin{figure}[h]
\caption{Nearest neighbour spin correlations $\langle S^{z}_{n}S^{z}_{n+1}
\rangle=\frac{1}{3} \langle {\mathbf S}_{n}{\mathbf S}_{n+1} \rangle$ in
the variational ground state of (\protect\ref{hamil}) for $\delta=0$ as
functions of $\alpha$ calculated from (\protect\ref{vark}) for different
system sizes $N$.}
\begin{center}
\epsfxsize=\textwidth
\epsffile{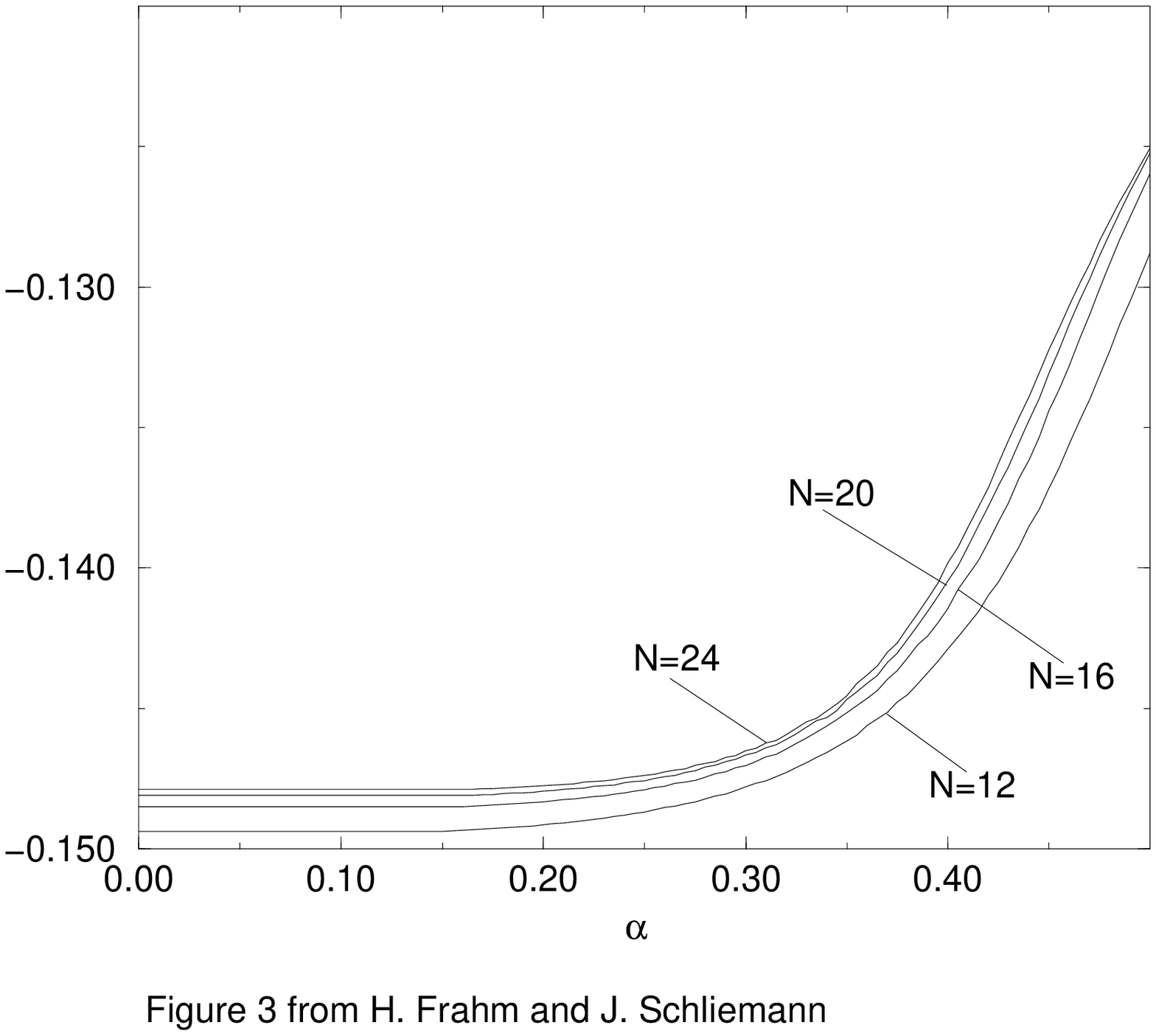}
\label{cor1}
\end{center}
\end{figure}

\begin{figure}
\caption{Next nearest neignbour spin correlations $\langle
S^{z}_{n}S^{z}_{n+2} \rangle=\frac{1}{3} \langle {\mathbf S}_{n}{\mathbf
S}_{n+2} \rangle$ of the variational ground state of (\protect\ref{hamil})
for $\delta=0$ as functions of $\alpha$ calculated from
(\protect\ref{vark}) for different system sizes $N$.}
\label{cor2}
\begin{center}
\epsfxsize=\textwidth
\epsffile{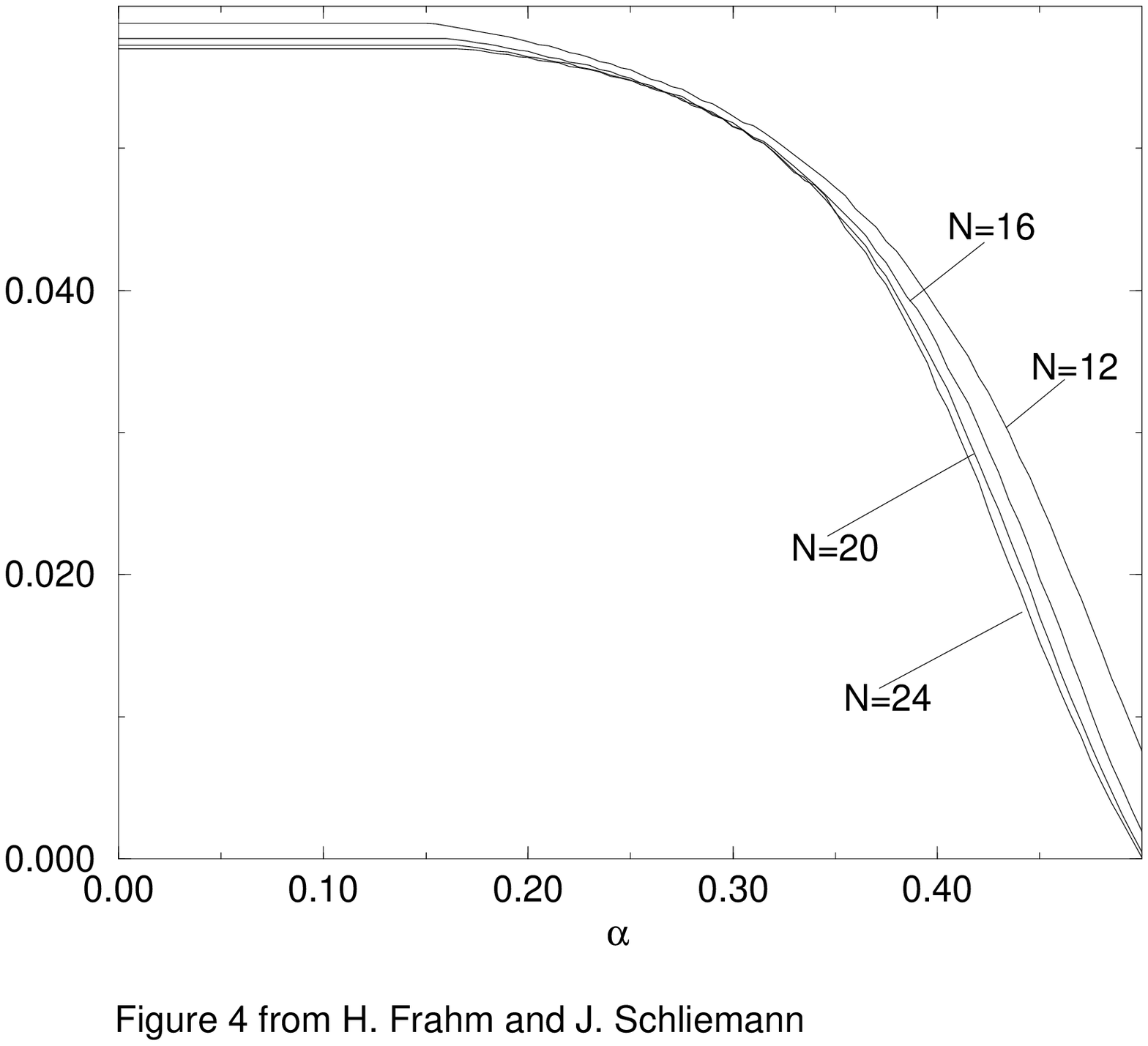}
\end{center}
\end{figure}

For the ground state of the model we have used the singlet-DJS ($M=J=N/2$)
as a variational ansatz.  The numerical results can be summarized as follows:

For $2\alpha+\delta\leq 1$ the expectation value of the hamiltonian is
minimized by real $w$, i.e.\ $\theta=\pi/2$ with $\rho$ varying from $0$ to
$1/2$, $2\alpha+\delta=1$ corresponds to a valence bond state ($w=1/2$)
\cite{shsu:81}.  On the other hand, for $2\alpha+\delta\geq 1$ we find
$\rho=1/2$ and varying $\theta$.  For $\alpha=\delta=0$ the minimum is
given by the Haldane--Shastry ground state ($w=0$).  The exact ground state
energy per spin is known to be $-\ln2+1/4\simeq-0.443147$ for an infinite
system, the variational value is $-(3/4){\rm Si}(\pi)/(\pi)
\simeq-0.442177$ from (\ref{CFHS}). In Table \ref{tab1} we present the
extrapolation to an infinite system of the variational ground state
energies for various values of $\alpha$ and $\delta$.  Comparison with data
from numerical diagonalisation obtained by Chitra et al.\ \cite{chiet:95}
shows excellent agreement: the variational energies per spin differ by only
about $10^{-3}$ from diagonalisation values except for the last row of
table \ref{tab1}.  In general, the quality of the singlet-DJS as a
variational ansatz is found to decrease for values of $\alpha$ exceeding
$0.5$.  
Hence, we have found an effectively one-parametric variational wave
function that is an excellent approximation of the true ground state of the
model considered within a large area of its parameters, in particular for
any antiferromagnetic nnn couplings $\alpha$ with $2\alpha+\delta\leq 1$.
Note that in this case the optimum singlet DJS (\ref{DJS}) has {\it real}
amplitudes.  For $2\alpha+\delta>1$ they become properly complex: this case
therefore is called a chiral phase. \\
For $\delta\neq 0$ the translational invariance of the system is explicitly
broken leading to a dimerized variational ground state with finite
$\rho$.\\
For $\delta=0$ and small values of the nnn coupling
$0\leq\alpha\leq\alpha^{\ast}$ the variational energy is minimized for
$w=0$.  A transition to a dimerized ground state is observed at the
`critical' value of $\alpha^{\ast}\simeq 0.2716\pm 0.0002$ for
$N\to\infty$. For $\alpha>\alpha^{\ast}$ $\rho$ becomes finite. In absence
of the alternating term the Hamiltonian ($\delta=0$) the system is
invariant under translation by a single lattice site, while the states
(\ref{DJS}) do not have such a symmetry for $w\neq 0$.  This leads to
consider the ansatz
\begin{equation}
  |\,\psi^{N}(\rho)\rangle\propto
  |\,\psi^{N,\frac{N}{2},\frac{N}{2}}(\frac{\pi}{2},\rho)\rangle\
  +|\,\psi^{N,\frac{N}{2},\frac{N}{2}}(\frac{\pi}{2},-\rho)\rangle
\label{vark}
\end{equation}
with $N$ chosen even.  This construction corresponds to the lattice
momentum of the ground state of finite systems obtained in
Ref.~\onlinecite{toha:87}.  With the states (\ref{vark}) one observes
slight, but numerically significant improvements of the ground state energy
as shown in table \ref{tab2}.  Furthermore, the parameter $\rho$ remains
zero for $0\leq\alpha\leq\alpha^{\ast\ast}$ with
$\alpha^{\ast\ast}\simeq0.1737\pm 0.0002$, which is much smaller than the
$\alpha^{\ast}$ mentioned above.
In Figure \ref{gs} we present the ground state energy per spin as a
function of $\alpha$, in Figures \ref{cor1},\ref{cor2} nn and nnn
correlations calculated within the states (\ref{vark}) are plotted. These
diagrams agree very well with the corresponding figures given in
Ref.~\onlinecite{toha:87}.

Spin-spin correlations beyond those entering the expression for the ground
state energy show the correct long distance asymptotics as long as $\rho=0$
(\ref{CFHS}), which corresponds to the massless regime $\alpha \lesssim
{1\over4}$, $\delta=0$ described by a conformal field theory.  We find that
non-zero $\rho$ lead to a suppression of long-range correlations.  The
system sizes that we have analyzised numerically do not allow, however, to
study the dependence of the correlation length on $\rho$.
%
%
%
\subsection{Excitations}
\begin{figure}
\caption{Energy gap between the variational ground state and the lowest
triplet $\Delta_{st}$ and the singlet (\protect\ref{sing})
$\Delta_{ss^{\ast}}$ for $\delta=0$ as a function of $\alpha$ for $N=20$.
The degeneracy determines our estimate of the conformal point $\alpha_c$.}
\label{excit}
\begin{center}
\epsfxsize=\textwidth
\epsffile{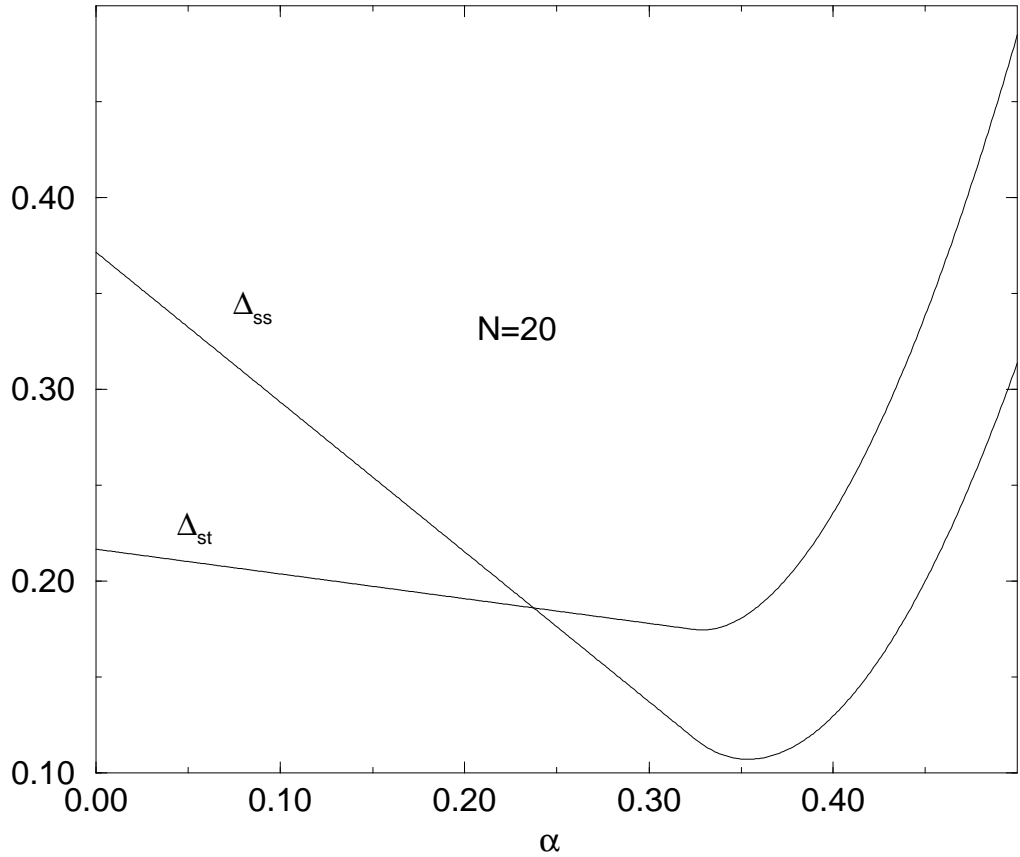}
\end{center}
\end{figure}

The results of the previous section suggest to use DJS with higher spin as
variational ansatz for excitations of our model.  Unfortunately the
situation is not as clear as before. Here we concentrate on the case
$\delta=0$ and mention that eigenstates of the Haldane--Shastry model
provide a good description of low-lying states for $0\leq \alpha\leq
0.3$.\\
{}From analytical and numerical studies this model is known to be gapless
for $\alpha\leq \alpha_{c}$. As predicted by conformal field theory, at the
`conformal point' $\alpha_{c}$ is defined by the occurence of many
degeneracies in addition to the usual SU(2) symmetry.  Okamoto and Nomura
examined the spectrum of (\ref{hamil}) for $\delta=0$ in finite systems by
numerical diagonalisation \cite{okno:92}.  They found a linear dependence
of the energy of the ground state and the two first excitations (triplet,
singlet) on $\alpha$ in the above intervall.  {}From the condition, that the
two excitations should degenerate, they obtained a precise value for the
conformal point as $\alpha_{c}=0.2411\pm 0.0001$.  Within the concept of
DJS, there is only one singlet corresponding to the groundstate, but in the
HS case ($w=0$) an additional singlet can be derived from the triplet
excitation by using the Yangian symmetry of this model \cite{halx:92}. \\
The Yangian of the Haldane--Shastry model is generated by the total
SU(2)-spin or `level-0 operators' $Q_{0}^{\alpha}=\sum S_{n}^{\alpha}$ and
the `level-1-operators' $Q_{1}^{\alpha}=\sum_{m\neq n}
\cot(\frac{\pi}{N}(m-n)) \varepsilon^{\alpha\beta\gamma}
S_{m}^{\beta}S_{n}^{\gamma}$.  We do not repeat any details of this
symmetry algebra and its representations here. In the HS model the lowest
excitation for an even number of spins is given by the triplet-DJS
$|\psi_{HS}^{N,N/2-1,N/2}\rangle$. This state degenerates with a singlet
that can be obtained by applying $Q_{1}^{-}= Q_{1}^{x}-\imath Q_{1}^{y}$ on
$|\psi_{HS}^{N,N/2-1,N/2}\rangle$ and projecting onto the singlet space,
i.e.\
\begin{equation}
  |\,\psi_{HS}^{S^{\ast}}\rangle\propto(2-{\mathbf Q}_{0}^{2})Q_{1}^{-}
  |\,\psi^{N,\frac{N}{2}-1,\frac{N}{2}}_{HS}\rangle\ .
\label{sing}
\end{equation}
As mentioned before, the minimum expectation value of the hamiltonian on
the line $\delta=0$ is given by the HS ground state for
$\alpha\leq\alpha^{\ast}$. So we evaluated the hamiltonian also for the two
excited HS states above. Note that in this ansatz there is no variational
parameter included, because (\ref{sing}) is not defined for $w\ne 0$. In
Figure \ref{excit} the difference of these energies with the ground state
value given in Figure \ref{gs} are plotted.  For $\alpha\leq 0.3$ there is
a good agreement with the corresponding data of Ref.\
\onlinecite{okno:92}. At some $\alpha=\tilde\alpha$ the singlet and triplet
energies coincide.  Table \ref{tab3} shows values for $\tilde\alpha$ in
finite systems, which can be extrapolated to $N\to\infty$ smoothly giving
$\tilde\alpha=0.2368\pm 0.0002$. This is remarkably close to the value
$\alpha_{c}=0.2411\pm 0.0001$ obtained from numerical diagonalisation.
%
%
\section{Conclusions}
We have constructed a family of variational states which contain excellent
approximations to the ground state of the model (\ref{hamil}) for a large
range of its parameters.  The states are strictly spin--singlets and can be
characterized by a single complex parameter.  Moreover, in
Section~\ref{sec:app} we have demonstrated that for $\delta=0$ and
$0\leq\alpha\leq 0.3$ groundstate and lowest excitations of our model can
be desribed amazingly well by eigenstates of the HS model.\\ 
For $\delta\neq 0$ the variational ansatz reproduces the explicit
dimerisation of the model.\cite{DimerGS} For vanishing bond alternation
$\delta=0$ the ground state of the finite system has a well-defined lattice
momentum $k_{s}\in\{0,\pi\}$ \cite{toha:87} while the lowest excitations
(triplet, singlet) have $k_{t},k_{s^{\ast}}=\pi-k_{s}\pmod{2\pi}$. In the
thermodynamic limit the excitation gap of the triplet vanishes for
$\alpha\leq\alpha_{c}$, while for larger $\alpha$ the singlet degenerates
with the ground state of the finite system.  The momenta of these two
states differ by $\pi$.  Consequently, in this two dimensional ground state
space linear combinations exist that represent dimerized states. This is
expressed within our variational approach by the fact that even for {\it
finite} systems dimerized states (with $\rho\neq 0$) are good
approximations of the ground state. The corresponding translational
invariant states (\ref{vark}) lead to small improvements to the ground
state energy, but reproduce the spin correlations between nearest and next
nearest neighbours very well.

As mentioned above a generalization of this variational states to similar
systems away from half filling is straightforward.  This may allow for
similar studies of the Zn--doped compound {Cu$_{1-x}$Zn$_{x}$GeO$_3$}
\cite{Dope} when the Zn-sites are treated as static spin-0
sites in the spin chain.
\section*{Acknowledgements}
The authors thank F.\,H.\,L.\,E\ss{}ler and M.\,Takahashi for useful
discussions on this topic.  This work has been supported by the Deutsche
Forschungsgemeinschaft under Grant No.\ Fr~737/2--2.
\newpage
\appendix
\section{Off-diagonal matrix elements of the DJS}
\label{appa}
To compute the action of
$\sum_{n<m}(S_{n}^{+}S_{m}^{-}+S_{n}^{-}S_{m}^{+})$ on the DJS (\ref{DJS})
we define
\begin{equation}
  |\chi(\{n_{i}\})\rangle=\prod_{i}S_{n_{i}}^{-}
  |\uparrow\cdots\uparrow\rangle
\label{ps}
\end{equation}
with $|\{n_{i}\}|=M$.  Introducing $z=e^{\imath\frac{2\pi}{N}}$ and
$\mu_{n_{i}}=(z^{2w})^{(-1)^{n_{i}}}$ we find
\begin{eqnarray}
  &&\frac{\langle\chi(\{n_{i}\})\,|\,\sum_{n<m}
  (S_{n}^{+}S_{m}^{-}+S_{n}^{-}S_{m}^{+})\,\psi^{N,M,J}(\theta,\rho)\rangle}
  {\langle\chi(\{n_{i}\})\,|\,\psi^{N,M,J}(\theta,\rho)\rangle}
\nonumber\\
  &&\qquad=
  \sum_{n=1}^{\frac{N}{2}-1}z^{J2n} 
  \sum_{i}\Bigl(\prod_{i\not=j\atop|n_{i}-n_{j}|\in2{\rm N}}
  \frac{(z^{n_{i}+2n}-z^{n_{j}})(z^{-n_{i}-2n}-z^{-n_{j}})}
  {(z^{n_{i}}-z^{n_{j}})(z^{-n_{i}}-z^{-n_{j}})}
\nonumber \\
  &&\quad\quad\quad\quad\quad\qquad\qquad
  \times\prod_{i\not=j\atop|n_{i}-n_{j}|\notin2{\rm N}}
  \frac{\mu_{n_{i}}z^{n_{i}+2n}
  -z^{n_{j}}}{\mu_{n_{i}}z^{n_{i}}
  -z^{n_{j}}}\,
  \frac{{1\over\mu_{n_i}}z^{-n_{i}-2n}
  -z^{-n_{j}}}{{1\over\mu_{n_i}}z^{-n_{i}}
  -z^{-n_{j}}}\,\Bigr)
\nonumber\\
  &&\qquad+\sum_{n=0}^{\frac{N}{2}-1}z^{J(2n+1)}
  \sum_{i}\,\Bigl(\prod_{i\not=j\atop|n_{i}-n_{j}|\in2{\rm N}}
  \frac{{1\over\mu_{n_i}}z^{n_{i}+2n+1}-z^{n_{j}}}
  {z^{n_{i}}-z^{n_{j}}}
  \frac{\mu_{n_{i}}z^{-n_{i}-2n-1}
  -z^{-n_{j}}}{z^{-n_{i}}-z^{-n_{j}}}
\nonumber\\
  &&\quad\quad\quad\quad\quad\qquad\qquad
  \times\prod_{i\not=j\atop|n_{i}-n_{j}|\notin2{\rm N}}
  \frac{z^{n_{i}+2n+1}-z^{n_{j}}}
  {\mu_{n_{i}}z^{n_{i}}-z^{n_{j}}}\,
  \frac{z^{-n_{i}-2n-1}-z^{-n_{j}}}
  {{1\over\mu_{n_i}}z^{-n_{i}}
  -z^{-n_{j}}}\,\Bigr)
\label{appa:umf}
\end{eqnarray}
The two sums originate from hopping terms between spins separated by an
even and odd number of lattice sites respectively. Combining these terms
(\ref{appa:umf}) can be rewritten as
\begin{equation}
\frac{\langle\chi(\{n_{i}\})\,|\,\sum_{n<m}
  (S_{n}^{+}S_{m}^{-}+S_{n}^{-}S_{m}^{+})\,\psi^{N,M,J}(\theta,\rho)\rangle}
  {\langle\chi(\{n_{i}\})\,|\,\psi^{N,M,J}(\theta,\rho)\rangle} +M
  =\sum_{n=0}^{N-1}z^{nJ}P((\pm\,z)^{n},(\pm\,z)^{-n})
\label{appa:poly}
\end{equation}
where $P$ is a polynomial in its arguments with coefficients independent
on $n$.  Only powers $(\pm z)^{nk}$ with $-(M-1)\leq k \leq(M-1)$ arise
in the expansion (\ref{appa:poly}), so that for integer $J$ with
$M\leq J\leq N-M$ the sum taken over each term of $P$ vanishes leading to
\begin{equation}
  \sum_{n<m}(S_{n}^{+}S_{m}^{-}+S_{n}^{-}S_{m}^{+})\,
  |\psi^{N,M,J}(\theta,\rho)\rangle
   =-M\,|\psi^{N,M,J}(\theta,\rho)\rangle\,,
\label{hopp}
\end{equation}
\newpage

\begin{thebibliography}{10}

\bibitem[*]{email_hf}
email: frahm@itp.uni-hannover.de

\bibitem[\dag]{addr_js}
Address after February 1, 1997: Physikalisches Institut, Universit\"at
Bayreuth, D-95440 Bayreuth, Germany;\\
email: John.Schliemann@theo.phy.uni-bayreuth.de

\bibitem{hase:93}
M. Hase, I. Terasaki, and K. Uchinokura, Phys. Rev. Lett. {\bf 70},  3651
  (1993);
M. Hase, I. Terasaki, K. Uchinokura, M. Tokunaga, N. Miura, and H. Obara, Phys.
  Rev. B {\bf 48},  9161  (1993).

\bibitem{lorx:94}
J.~E. Lorenzo, K. Hirota, G. Shirane, J.~M. Tranquada, M. Hase, K. Uchinokura,
  H. Kojima, I. Tanaka, and Y. Shibuya, Phys. Rev. B {\bf 50},  1278  (1994).

\bibitem{cace:95}
G. Castilla, S. Chakravarty, and V.~J. Emery, Phys. Rev. Lett. {\bf 75},  1823
  (1995).

\bibitem{bethe:31}
H. Bethe, Z. Phys. {\bf 71},  205  (1931).

\bibitem{magh:69}
C.~K. Majumdar and D.~K. Ghosh, J. Math. Phys. {\bf 10},  1388  (1969).

\bibitem{AKLT}
I. Affleck, T. Kennedy, E.~H. Lieb, H. Tasaki, Commun. Math. Phys. 
  {\bf 115} 477 (1988).

\bibitem{affx:89}
I. Affleck, D. Gepner, H.~J. Schulz, and T. Ziman, J. Phys. A {\bf 22},  511
  (1989).

\bibitem{okno:92}
K. Okamoto and K. Nomura, Phys. Lett. A {\bf 169},  433  (1992).

\bibitem{whaf:96}
S.~R. White and I. Affleck, Phys. Rev. B {\bf 54},  9862  (1996).

\bibitem{DimerGS}
E.~H. Lieb and B. Nachtergaele, Phys. Rev. B, {\bf 51}, 4777 (1995);
M. Aizenman and B. Nachtergaele, Commun. Math. Phys., {\bf 164}, 17 (1994).

\bibitem{shsu:81}
B.~S. Shastry and B. Sutherland, Phys. Rev. Lett. {\bf 47},  964  (1981).

\bibitem{Ladders}
T. Barnes, E. Dagotto, J. Riera, and E.~S. Swanson, Phys. Rev. B {\bf 47},
  3196  (1993);
S. Gopalan, T.~M. Rice, and M. Sigrist, Phys. Rev. B {\bf 49},  8901  (1994).

\bibitem{MeanField}
M. Azzouz and C. Bourbonnais, Phys. Rev. B {\bf 53},  5090  (1996);
V.~N. Muthukumar, C. Gros, W. Wenzel, R. Valenti, P. Lemmens, B. Eisener, G.
  G{\"u}ntherodt, M. Weiden, C. Geibel, and F. Steglich, Phys. Rev. B {\bf 54},
   R9635  (1996);
V.~N. Muthukumar, C. Gros, R. Valenti, M. Weiden, C. Geibel, F. Steglich, P.
  Lemmens, M. Fischer, and G. G{\"u}ntherodt, preprint
  cond-mat/9611018.

\bibitem{kahf:82}
T.~A. Kaplan, P. Horsch, and P. Fulde, Phys. Rev. Lett. {\bf 49},  889  (1982).

\bibitem{HaldaneShastry}
F.~D.~M. Haldane, Phys. Rev. Lett. {\bf 60},  635  (1988);
B.~S. Shastry, Phys. Rev. Lett. {\bf 60},  639  (1988).

\bibitem{hald:91}
F.~D.~M. Haldane, Phys. Rev. Lett. {\bf 66},  1529  (1991).

\bibitem{ValenceBonds}
C. Zeng and J.~B. Parkinson, Phys. Rev. B {\bf 51},  11609  (1995);
S. Brehmer, H.-J. Mikeska, and U. Neugebauer, J. Phys. Condens. Matter {\bf 8},
   7161  (1996);
S. Yamamoto, Phys. Rev. B {\bf 55}, 3603 (1997);
D. Khomskii, W. Geertsma, and M. Mostovoy, preprint cond-mat/9609244.

\bibitem{wwz:89}
X.~G. Wen, F. Wilczek, and A. Zee, Phys. Rev. B {\bf 39}, 11413 (1989).

\bibitem{GeVo}
F. Gebhard and D. Vollhardt, Phys. Rev. Lett. {\bf 59},  1472  (1987);
   Phys. Rev. B {\bf 38},  6911  (1988).

\bibitem{nata:95}
H. Nakano and M. Takahashi, J. Phys. Soc. Japan {\bf 64},  2762  (1995).

\bibitem{gutz:63}
M.~C. Gutzwiller, Phys. Rev. Lett. {\bf 10},  159  (1963).

\bibitem{MeVo}
W. Metzner and D. Vollhardt, Phys. Rev. Lett. {\bf 59},  121  (1987);
   Phys. Rev. B {\bf 37},  7382  (1988).

\bibitem{tJJast}
Y. Kuramoto and H. Yokoyama, Phys. Rev. Lett. {\bf 67},  1338  (1991);
H. Yokoyama and M. Ogata, Phys. Rev. Lett. {\bf 67},  3610  (1991).

\bibitem{schliemann:dipl}
J. Schliemann, Diploma thesis, Universit{\"a}t Hannover, 1996.

\bibitem{chiet:95}
R. Chitra, S. Pati, , H.~R. Krishnamurthy, D. Sen, and S. Ramasesha, Phys. Rev.
  B {\bf 52},  6581  (1995).

\bibitem{toha:87}
T. Tonegawa and I. Harada, J. Phys. Soc. Japan {\bf 56},  2153  (1987).

\bibitem{halx:92}
F.~D.~M. Haldane, Z.~N.~C. Ha, J.~C. Talstra, D. Bernard, and V. Pasquier,
  Phys. Rev. Lett. {\bf 69},  2021  (1992).

\bibitem{Dope}
M. Hase, I. Terasaki, Y. Sasago, K. Uchinokura, and H. Obara, Phys. Rev. Lett.
  {\bf 71},  4059  (1993);
Y. Sasago, N. Koide, K. Uchinokura, M.~C. Martin, M. Hase, K. Hirota, and G.
  Shirane, Phys. Rev. B {\bf 54},  R6835  (1996).

\end{thebibliography}

\newpage

\begin{table}
\caption{Variational ground state energies per spin for $N\to\infty$.  The
last column shows numerical data from Ref.~\protect\onlinecite{chiet:95}.}
\label{tab1}
\begin{tabular}{cccc}
$\alpha$ & $\delta$ & $E^{var}_{S}(\alpha,\delta,\infty)$
& $E^{diag}_{0}/N$\\
\hline
$0.2411$ & $0.00$ & $-0.40160$ & $-0.401866$ \\
$0.2411$ & $0.04$ & $-0.40793$ & $-0.409051$ \\
$0.2411$ & $0.16$ & $-0.44237$ & $-0.442862$ \\
$0.2411$ & $0.32$ & $-0.49673$ & $-0.496844$ \\
$0.2500$ & $0.00$ & $-0.40008$ & $-0.40045$  \\
$0.2500$ & $0.35$ & $-0.50721$ & $-0.50727$  \\
$0.2500$ & $0.20$ & $-0.46242$ & $-0.46329$  \\
\hline
$0.4000$ & $0.50$ & $-0.56550$ & $-0.56611$  \\
$0.4800$ & $0.10$ & $-0.41272$ & $-0.41281$  \\
$0.5500$ & $0.10$ & $-0.41519$ & $-0.41610$  \\
$0.2000$ & $0.80$ & $-0.67601$ & $-0.67613$  \\
$0.3000$ & $0.80$ & $-0.67896$ & $-0.67966$  \\
$0.4800$ & $0.80$ & $-0.68897$ & $-0.69256$  \\
\end{tabular}
\end{table}

\begin{table}
\caption{Ground state energies per spin at $\delta=0$ and $N\to\infty$.
The second column shows variational results obtained from (\ref{DJS}), the
third from (\ref{vark}). The data in the last column is taken again from
Ref.~\protect\onlinecite{chiet:95}.}
\label{tab2}
\begin{tabular}{cccc}
$\alpha$ & $E^{var}_{S}(\alpha,0,\infty)$ & 
	$E^{var}_{S,k=0}(\alpha,0,\infty)$ & $E^{diag}_{0}/N$ \\
\hline
$0.20$ & $-0.40852$ & $-0.40880$ & $-0.40885$ \\
$0.25$ & $-0.40008$ & $-0.40045$ & $-0.40045$ \\
$0.30$ & $-0.39158$ & $-0.39240$ & $-0.39284$ \\
$0.40$ & $-0.37919$ & $-0.37924$ & $-0.38028$ \\
\end{tabular}
\end{table}

\begin{table}
\caption{Location of the degeneracy $\tilde\alpha$ between singlet and
triplet excitation for different $N$}
\label{tab3}
\begin{tabular}{rl}
$N$ & $\tilde{\alpha}$ \\
\hline
$10$ & $0.23967$ \\
$12$ & $0.23876$ \\
$14$ & $0.23820$ \\
$16$ & $0.23785$ \\
$18$ & $0.23757$ \\
$20$ & $0.23739$ \\
$22$ & $0.23722$ \\
\hline
$\infty$ & $0.2368$ \\
\end{tabular}
\end{table}

\newpage

\newpage

\newpage

\newpage

\newpage

\end{document}